\documentclass[cup6b]{cupbook}
\usepackage{graphicx}
\usepackage{natbib}
\title[Rome, Italy, 27--30 April 2009]
      {The coming of age of X-ray polarimetry}
\author{}
\date{}
\begin{document}
\pagenumbering{arabic}

\author[Vladim\'ir Karas]
{Vladim\'ir Karas (Astronomical Institute, Academy of Sciences, Prague)}

\chapter{Strong-gravity effects acting on polarization from orbiting spots}

\abstract{Accretion onto black holes often proceeds via an accretion disc or a
temporary disc-like pattern. Variability features, observed in the light
curves of such objects, and theoretical models of accretion flows suggest that
accretion discs are inhomogeneous and non-axisymmetric. Fast orbital motion of the
individual clumps can modulate the observed signal. If the emission from
these clumps is partially polarized, which is likely the case, then
rapid polarization changes of the observed signal are expected as a
result of general relativity effects.}

\section{Introduction}
Polarization of light originating from different regions of a black hole
accretion disc and detected by a distant observer is influenced by strong
gravitational field near a central black hole. A `spotted' accretion disc 
is a useful model of an interface of such an inhomogeneous medium, assuming 
that there is a well defined boundary between the disc
interior and the outer, relatively empty but highly curved spacetime. 
Relativistic corrections
to a signal from orbiting spots can lead to large rotation in the
plane of observed X-ray polarization. When integrated over an extended
surface of the source, this can diminish the observed degree of polarization. Such
effects are potentially observable and can be used to distinguish among
different models of the source geometry and the radiation mechanisms
responsible for the origin of the polarized signal. The polarization
features show specific energy and time dependencies which can indicate
whether a black hole is present in a compact X-ray source.

Practical implementation of the idea, originally proposed in the late 
1970s by Connors et al.\ \cite{con77,con80} and Pinneault 
\cite{pin77}, is a challenging task because the
polarimetric investigations need a high signal-to-noise ratio.
Also, the interpretation of the model results is often very sensitive to
the assumptions about the radiation transfer in the source and the
geometrical shape and orientation of the emission region. Nevertheless,
the technology has achieved significant advances since the 1980s and
reached a mature state, as demonstrated also by this Volume. Likewise,
numerous theoretical papers have made progress in our understanding of
the effects that we should look for.
We assume that the gravitational field is described by a rotating black 
hole, and so the Kerr metric is the right model for the gravitational field. 

On the whole, there are some similarities as well as differences between
the expected manifestation of GR polarization changes in X-rays and in
other spectral bands, such as the infrared region. We will mention these
interrelations and point out that the near-infrared polarization
measurements of the radiation flares from the immediate vicinity of the
horizon are already now available for Sagittarius A* supermassive black
hole in the Galactic Center \cite{mey06,zam08}.

\section{Time-varying polarization from orbiting spots}

The model of an orbiting bright spot \cite{bro05,cun72,mey06,nob07} 
has been fairly successful in
explaining the observed modulation of various accreting black hole
sources. Certainly not all variability patterns can be explained in
this way, however, the scheme is general enough to be able to capture
also the effects of spiral waves and similar kind of transient phenomena
that are expected to occur in the disc \cite{kar01,tag90,tag06}.
It can be argued that the spot lightcurves can be phenomenologically
understood as a region of enhanced emission that performs a
co-rotational motion near above the innermost stable circular orbit
(ISCO). For example, within the framework of the flare--spot model
\cite{cze04} the spots are just regions of enhanced emission on 
the disc surface rather than massive clumps that could suffer from fast
decay due to shearing motion in the disc. The observed signal is
modulated by relativistic effects. According to this idea, Doppler and
gravitational lensing influence the observed radiation flux and this can
be computed by ray-tracing methods. Such an approach has been extended
to compute also strong gravity effects acting on polarization properties
\cite{dov04c}.

A spot on the disc surface is supposed to be 
intrinsically polarized (by different possible mechanisms -- either
by reflection of a primary flare on the disc surface or by synchrotron
emission originating from an expanding blob, as detailed below). It
represents a rotating surface feature which shares the bulk orbital
motion of the underlying medium at sufficiently large radii above the
ISCO, gradually decaying due to differential rotation of the disc. 

We have applied different prescriptions for the local polarization
(see ref.\ \cite{dov06,mey06,zam08} for the detailed description of the model set-up in the
individual cases that we investigated). For example,
one set of models assumes the local emission to be polarized either in
the direction normal to the disc plane, or perpendicular to the toroidal
magnetic field. Obviously, in the case of partial local polarization
the observed polarization signal will be diluted by an unpolarized 
fraction, and so the polarization degree of the final signal will be 
proportionally diminished. In another set of models 
we assumed a lamp-post illuminated spot as the source of spot polarization 
by reflection. For the spot shape we first assumed the spot does not change
its shape during its orbit, but then we also consider the spot decay with
time. The relativistic effects can be clearly identified and understood with
these simple (and astrophysically unrealistic) toy models, as they
produce visible signatures in the observed polarization properties.

\begin{figure}
\centering
\includegraphics[width=0.49\textwidth]{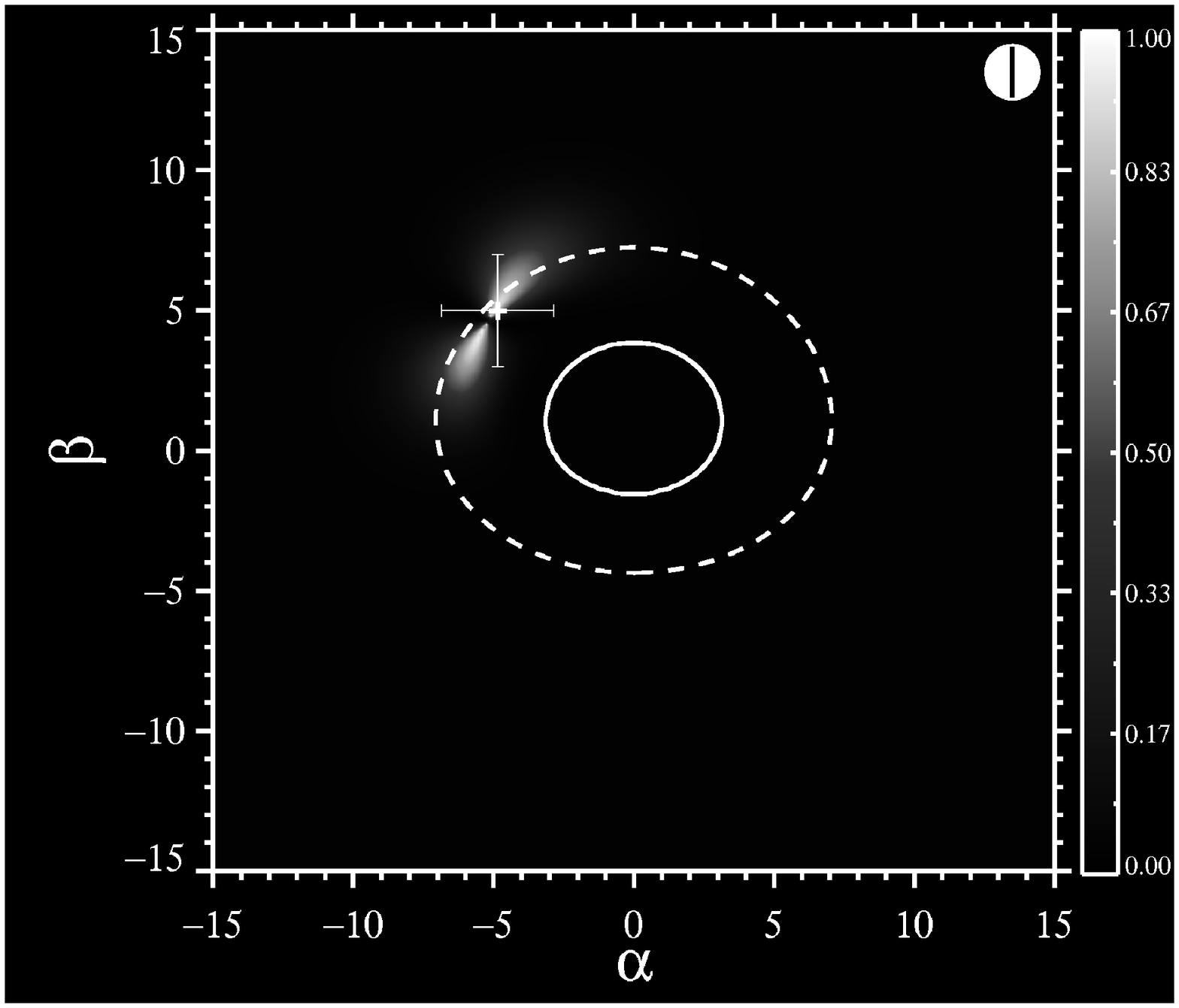}
\hfill
\includegraphics[width=0.49\textwidth]{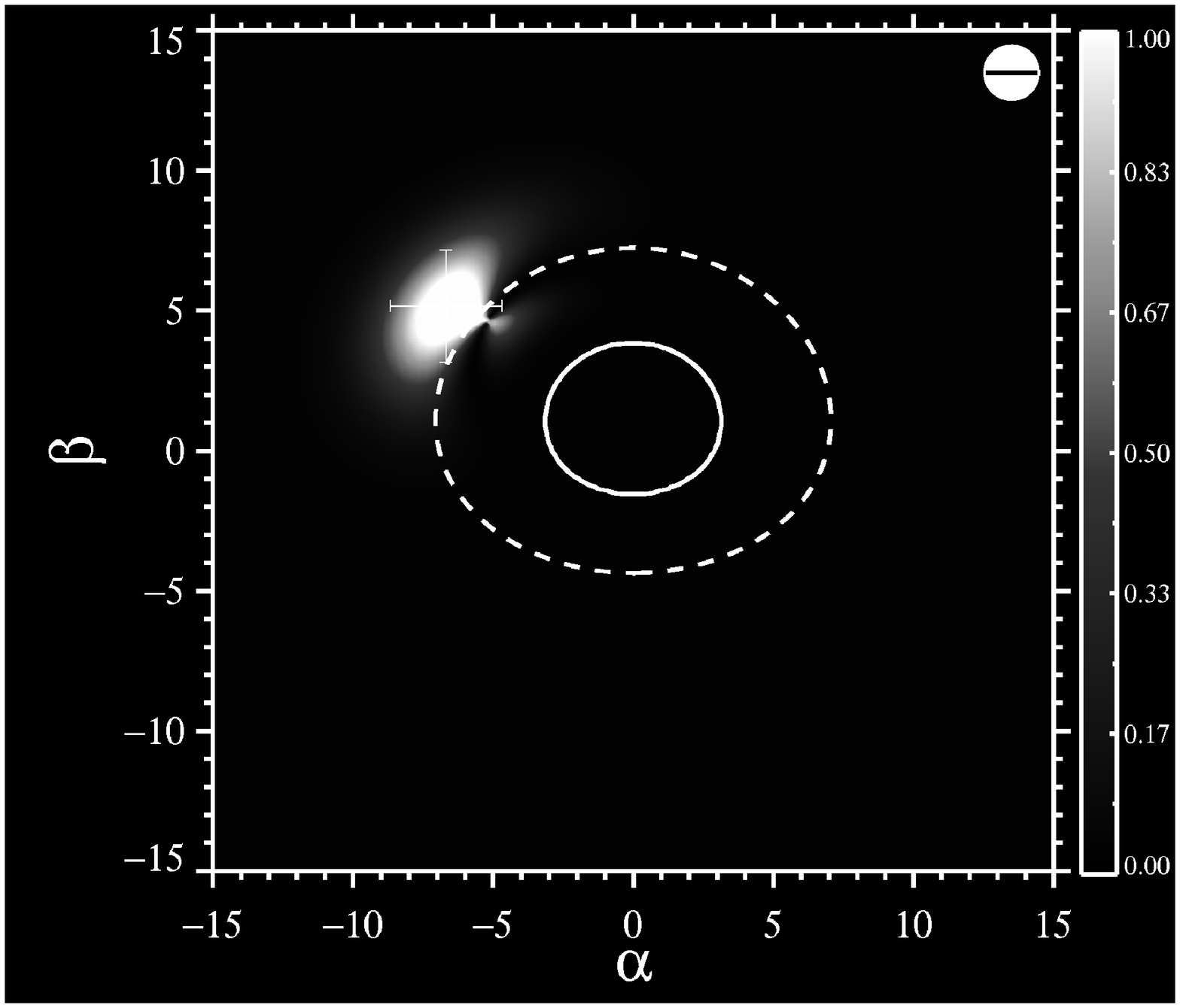}
\caption{A snapshot of a spot orbiting at constant radius $r=1.1r_{_{\rm ISCO}}$.
The image is produced by the spot emission that is assumed to be
intrinsically polarized and recorded in two polarization channels, 
rotated by 90 degrees with respect to each other \cite{zam09}.
The direction of the polarization filter is indicated in the top
right corner of each of the two panels.
The image is shown in the observer plane ($\alpha,\beta$), for a
non-rotating black hole observed at a moderate view angle, 
$\theta_{\rm{}o}=45$~deg. The horizon radius (solid curve) and 
the ISCO (dashed curve) are also shown.}
\label{fig1}
\end{figure}

General relativistic effects present in our model can be split into two
categories. Firstly, it is the symmetry breaking between the approaching
and the receding part of the spot orbit. Doppler beaming as well as the
light focusing contribute to the change of the observed flux, especially
at high view angles when the spot orbit is seen almost edge-on. Notice
that the Doppler boosting effect is off phase with respect to the light
focusing effect, roughly by 0.25 of the full orbit at the corresponding
radius. Here, the precise number depends on the black hole spin; it also
depends on the inclination through the finite light-travel time from
different parts of the spot orbit towards the observer. Also, higher
order images could be important in case of almost edge-on view of the
spot.

Secondly, rotation of the polarization plane along the photon trajectory
also plays a role. This effect is particularly strong for
small radii of the spot orbit, in which case a critical point occurs
\cite{dov08}. The observed polarization angle exhibits just a small
wobbling around its principal direction when the spot radius is above
the critical point, whereas it starts turning around the full circle
once the radius drops below the critical one. Notice that the exact
location of the critical point depends on the black hole angular
momentum, in principle allowing us to determine its value. 

However, a caveat (and a third point on the list) is caused by
sensitivity of the critical radius to the special relativistic
aberration effects, especially at small view angles (i.e. when the spot
is seen almost along the rotation axis). This means that the moment when
the observed polarization angle starts rotating is sensitive to the
underlying assumption of a perfectly planar geometry of the disc
surface. 

\begin{figure}
\centering
\includegraphics[width=0.49\textwidth]{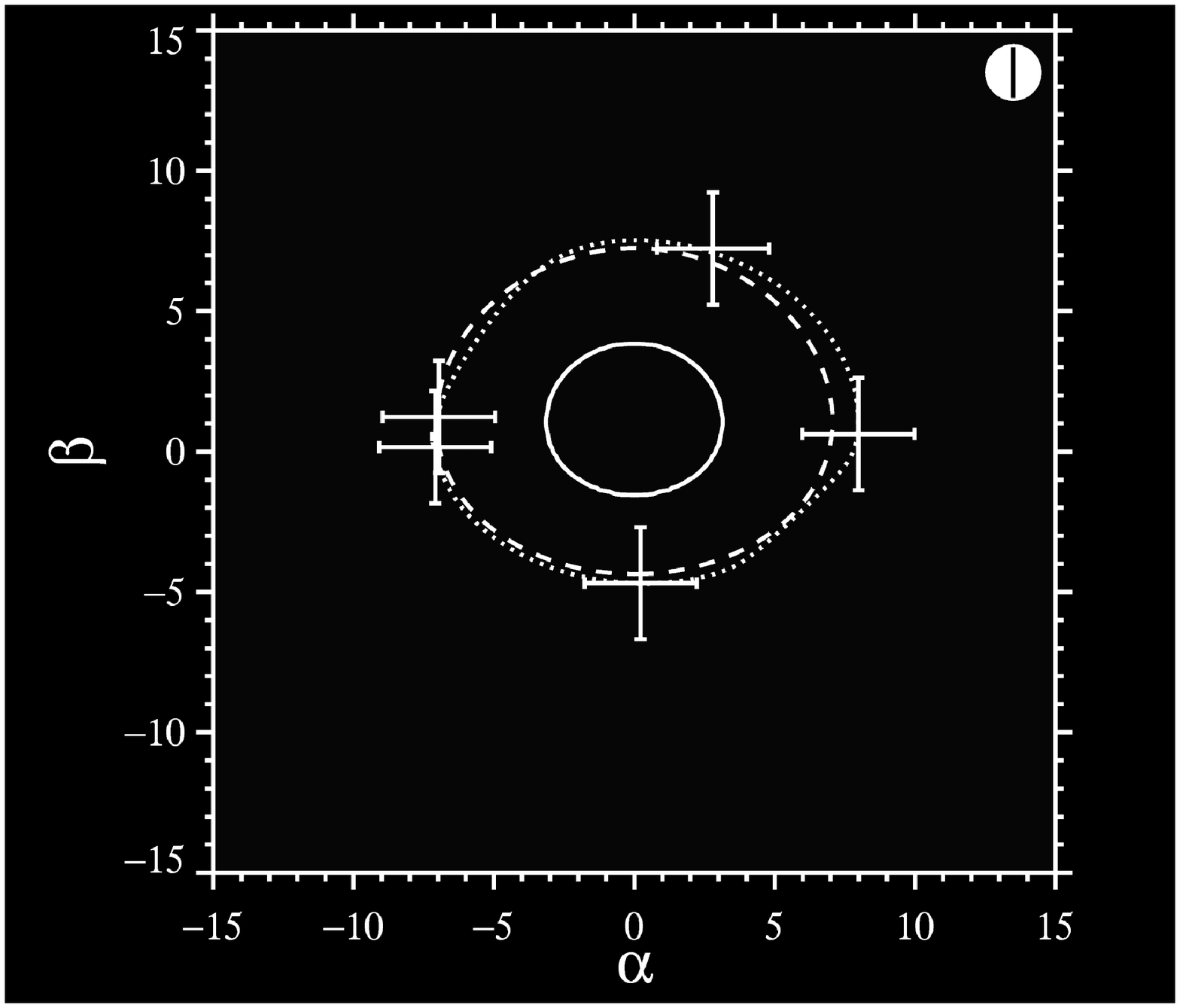}
\hfill
\includegraphics[width=0.49\textwidth]{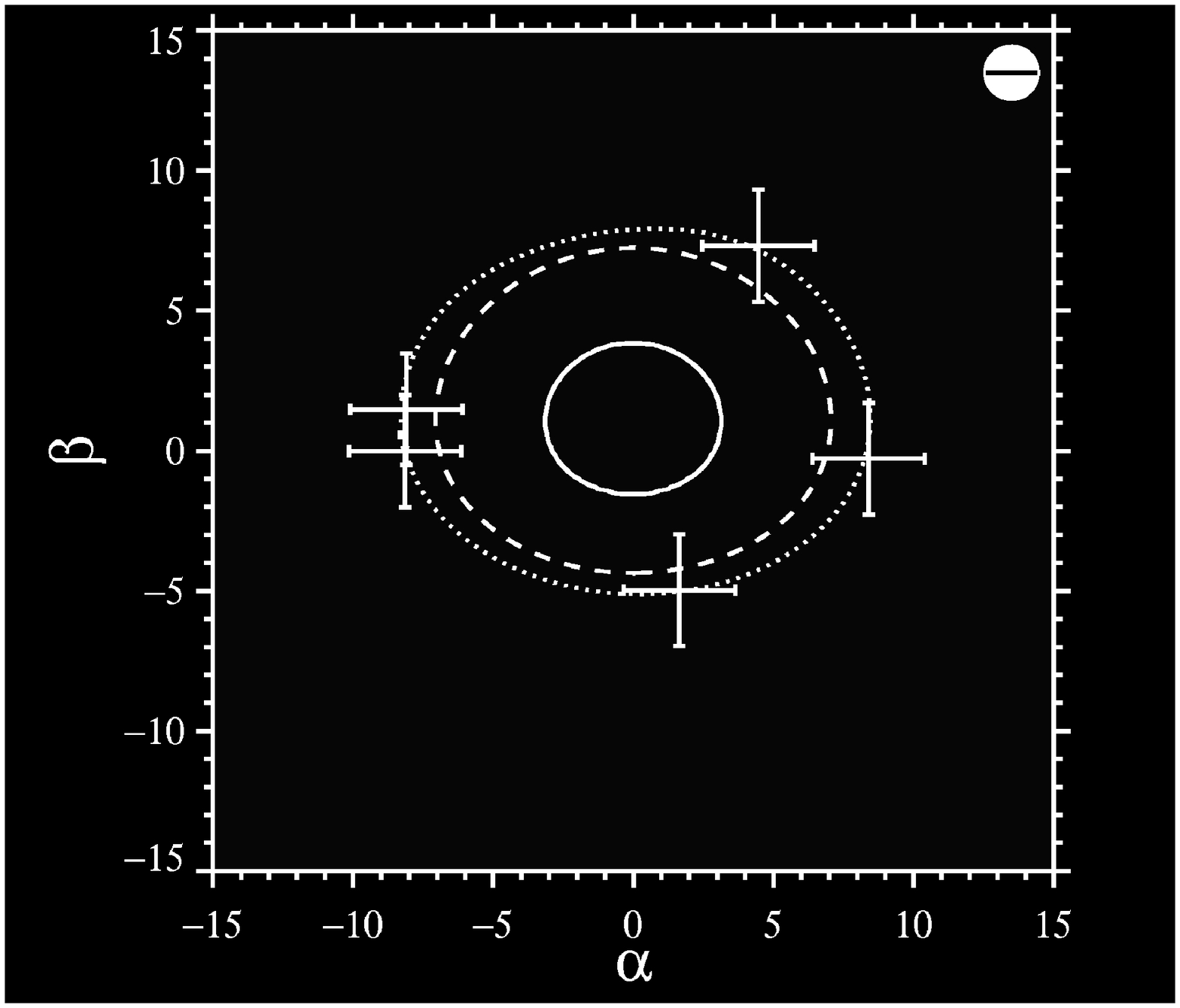}
\caption{Trajectory of the image centroid during one revolution of the spot
corresponding to the previous figure. The wobbling position of the image
centroid is indicated by crosses at five different moments along the image
track (dotted curve). Figures are courtesy of M.~Zamaninasab who has applied
the spot model to investigate the polarization properties of near-infrared 
flares from the Galactic Center supermassive black hole \cite{zam09}.}
\label{fig2}
\end{figure}

By combining the above-mentioned effects together, Dov\v{c}iak et al.\ \cite{dov06} have
shown that the observed polarization degree is expected to decrease (in
all their models) mainly in that part of the orbit where the spot moves
close to the region where the photons are emitted perpendicularly to the
disc. In this situation the polarization angle changes rapidly. The
decrease in the observed polarization degree for the local polarization
perpendicular to the toroidal magnetic field happens also in those parts
of the orbit where the magnetic field points approximately along light
ray.

For the more realistic models the resulting polarization shows
a much more complex behaviour. Among persisting features is the peak
in polarization degree for the extreme Kerr black hole for large
inclinations, caused by the lensing effect at a particular position of
the spot in the orbit where the  polarization angle is changed. This is
not visible in the Schwarzschild case.

The X-ray polarization lightcurves and spectra are still to be taken by
future missions, but one may envision even a more challenging goal
connected with imaging of the inner regions of accreting black hole
sources. Obviously this is a truly distant future: imaging a black hole
shadow would require order of ten microsecond angular resolution.
However, what might be realistically foreseen is the tracking of the
wobbling image centroid that a spot is supposed to produce
\cite{ham09,zam09}. With the polarimetric resolution, the wobbling could
provide an excellent evidence proving the presence of the orbiting
feature. See Fig.~\ref{fig1} for an example of the expected form of
the spot images and the corresponding centroid tracks in a simplified case 
of a model spot endowed with an
intrinsic polarization that remains constant in the co-orbiting 
frame.
This example assumes a spot rotating rigidly at constant radius near above the
ISCO. Orientation of the polarization filter is also fixed, as indicated
in the top-right corner of the plot. Correspondingly, Fig.~\ref{fig2} shows 
the tracks  of the image centroid. Albeit the tracks are not identical in the two
orientations of the polarization filter, the difference is rather subtle.

Notice that the project of detecting the centroid motion does not
necessarily have to be limited to the X-ray domain. In view of
recent results on Sagittarius A* flares, which have been reported in
X-rays as well as in the near infrared, submillimiter and the radio spectral bands
\cite{bag01,eck08,gen03,por03}, the immediate vicinity of the black hole can be
probed by various techniques. The simultaneous time-dependent
measurements equipped with the polarimetric resolution seem to be a
final goal of this effort.

\section{Conclusions}\label{sec:concl}
Polarimetry is know to be a photon-hungry technique, and so 
it is not easy to identify the
specific effects of general relativity that could be observed with
available polarimeters (in whatever spectral band) or with those
envisaged for realistically foreseeable future. In several recent
papers, and in particular in this Volume, various people demonstrate
that the task of detecting the relativistic effects and in this way
determining the physical parameters of the black hole systems seems to
be feasible. Among ways to reach the goal, time-dependent polarization
profiles, such as those expected from orbiting spots, play an important
role.

It may be worth reminding the reader that the KY code, employed
in our computations, is publicly available, either as a part of the XSPEC
package or directly from the authors \cite{dov04c}. The current version allows
the user to include the polarimetric resolution and to compute the observational
consequences of strong-gravity effects from a Kerr black hole accretion disc.
Within the XSPEC notation, this polarimetric resolution is encoded by a switch 
defining which of the four Stokes parameters is returned in the photon count array
at the moment of the output from the model evaluation. This way one can test 
and combine various models, and pass the resulting signal through the response 
matrices of different instruments.

\smallskip

Mohammad Zamaninasab kindly created figures for this article.
The author thanks the Czech Science Foundation (ref. 202/09/0772) and
the Center for Theoretical Astrophysics in Prague (ref. LC06014) for 
support.


\begin{thereferences}{99}
\bibitem{bag01}Baganoff F.~K., Bautz M.~W., Brandt W.~N., et al.\ (2001), \textit{Nature}, \textbf{413}, 45
\bibitem{bro05}Broderick A.~E., Loeb~A. (2005), \textit{MNRAS}, \textbf{367}, 905
\bibitem{con77}Connors P.~A., Stark R.~F. (1977), \textit{Nature} \textbf{269}, 128
\bibitem{con80}Connors P.~A., Stark R.~F., Piran~T. (1980), \textit{ApJ}, \textbf{235}, 224
\bibitem{cun72}Cunningham C.~T., Bardeen J.~M. (1972), \textit{ApJ}, \textbf{173}, L137
\bibitem{cze04}Czerny~B., R\'o\.za\'nska~A., Dov\v{c}iak~M., et al.\  (2004), A\&A, \textbf{420}, 1
\bibitem{dov06}Dov\v{c}iak~M., Karas~V., Matt~G. (2006), \textit{AN}, \textbf{327}, 993
\bibitem{dov04c}Dov\v{c}iak M., Karas V., Yaqoob T. (2004), \textit{ApJS}, \textbf{153}, 205
\bibitem{dov08}Dov\v{c}iak~M., Muleri~F., Goosmann R.~W., et al.\ (2008), \textit{MNRAS}, \textbf{391}, 32
\bibitem{eck08}Eckart~A., Baganoff F.~K., Zamaninasab~M., et al.\ (2008), \textit{A\&A}, \textbf{479}, 625
\bibitem{gen03}Genzel~R., Sch\"odel~R., Ott~T., et al.\ (2003), \textit{Nature}, \textbf{425}, 934
\bibitem{ham09}Hamaus~N., Paumard~T., M\"uller~T., et al.\ (2009), \textit{ApJ}, \textbf{692}, 902
\bibitem{kar01}Karas~V., Martocchia~A., \v{S}ubr~L. (2001), \textit{PASJ}, \textbf{53},~189
\bibitem{mey06}Meyer~L., Eckart~A., Sch\"odel~R., et al.\ (2006), \textit{A\&A}, \textbf{460}, 15
\bibitem{nob07}Noble S.~C., Leung Po Kin, Gammie C.~F., Book L.~G. (2007), \textit{CQG}, \textbf{24}, S259
\bibitem{pin77}Pineault~S. (1977), \textit{MNRAS}, \textbf{179}, 691
\bibitem{por03}Porquet~D., Predehl~P.. Aschenbach~B., et al.\ \textit{A\&A}, \textbf{407}, L17
\bibitem{tag90}Tagger~M., Henriksen R.~N., Sygnet J.~F., Pellat~R. (1990), \textit{ApJ}, \textbf{353}, 654
\bibitem{tag06}Tagger~M., Melia~F. (2006), \textit{ApJ}, \textbf{636}, 33
\bibitem{zam08}Zamaninasab~M., Eckart~A., Meyer~L., et al.\ (2008), \textit{J. Phys.}, \textbf{131}, 012008
\bibitem{zam09}Zamaninasab~M., Eckart~A., Witzel~G., et al.\ (2009), A\&A, submitted
\end{thereferences}
\end{document}